\documentstyle[11pt,twoside]{article}
 \newcommand{\RR}{{\bf R}^{1+s}}
 \newcommand{\N}{{\bf N}}
 \newcommand{\df}{\mbox{\,{\rm :}$=$\,}}
 \newcommand{\Bix}{\rule{0.6em}{0.6em}}
 \newcommand{\vlk}{{\bf V}_{\!\!+}}
 \newcommand{\cA}{{\cal{A}}}
 \newcommand{\cB}{{\cal{B}}}
 \newcommand{\cC}{{\cal{C}}}
 \newcommand{\cD}{{\cal{D}}}
 \newcommand{\cH}{{\cal{H}}}
 \newcommand{\cI}{{\cal{I}}}
 \newcommand{\cK}{{\cal{K}}}
 \newcommand{\cN}{{\cal{N}}}
 \newcommand{\cO}{{\cal{O}}}
 \newcommand{\cS}{{\cal{S}}}
 \newcommand{\cU}{{\cal{U}}}
 \newtheorem{prop}{Proposition}[section]
 \newtheorem{lem}[prop]{Lemma}

\title{On Infravacua and Superselection Theory}
\author{Walter Kunhardt
     \\ Institut f\"ur Theoretische Physik der Universit\"at G\"ottingen
     \\ Bunsenstra\ss{}e 9, 37073 G\"ottingen, Germany
     \\ e-mail: {\tt kunhardt@theorie.physik.uni-goettingen.de} }
\date{April 29, 1997}

\begin{document}

\maketitle

\begin{abstract}
In the DHR theory of superselection sectors, one usually considers states
which are local excitations of some vacuum state. Here, we extend this
analysis to local excitations of a class of  ``infravacuum'' states
appearing in models with massless particles. We show that the
corresponding superselection structure, the statistics of
superselection sectors and the energy-momentum spectrum are the same as with
respect to the vacuum state. (The latter result is
obtained with a novel method of expressing the shape of the spectrum
in terms of properties of local charge transfer cocycles.)
These findings provide evidence to the effect that infravacua are a
natural starting point for the analysis of the superselection structure
in theories with long-range forces.
\end{abstract}

\section{Introduction}
The description of the superselection structure of theories with
long-range interactions such as QED is a
long-standing problem  in the algebraic approach to quantum field theory. As
a characteristic feature, such theories contain charges which obey Gauss' law
and thus do not fit into the framework of superselection sectors as described
by Doplicher, Haag and Roberts \cite{DHR}.

Focusing (mainly for definiteness) on QED, Gauss' law has the following three
consequences: First, since every charge is accompanied by an electromagnetic
field, an operator $\psi_c$ which describes the creation of an electric charge
in a gauge-invariant way in the ``physical'' Hilbert space, cannot be local.
There is, however, the possibility of localising the charge and (the Cauchy
data of) its field in a space-like cone $\cC$ (as a simple example of a region
extending to space-like infinity). In this case, the charge cannot be
detected by measurements in the space-like complement $\cC'$ of that cone.

Second, the configuration of the asymptotic electric flux at space-like
infinity is a classical observable whose values label an uncountable number
of superselection sectors in the physical Hilbert space. Among these is the
vacuum sector, but also the sector generated from the vacuum by the operator
$\psi_c$. The former is identified with the vacuum representation $\pi_0$  of
the observables, the latter with the representation $\pi_0\circ\gamma_{\cC}$,
where $\gamma_{\cC}\equiv{\rm Ad}\psi_c^*$. Obviously,  the space-like
direction of
$\cC $ can be determined by measurements of the asymptotic flux distribution
in the sector $\pi_0\circ\gamma_{\cC}$. This implies that the representations
$\pi_0$ and   $\pi_0\circ\gamma_{\cC}$, equal by construction when restricted
to
$\cC' $, will not be equivalent in restriction to the space-like complement
$\cC_1'$ of some arbitrary space-like cone $\cC_1$.

Third, Gauss' law lies at the origin of what is called the infraparticle
problem: electrons are infraparticles, i.e.\ they do not correspond to
eigenvalues of the mass operator. The physical picture for this is that every
electron is accompanied by a cloud of low-energy photons. Related to this is
the emission of bremsstrahlung in collisions of charged particles. The creation
of infinitely many photons in such a process manifests itself on the
mathematical side in the fact that the representations of the incoming and the
outgoing free photon fields cannot both be Fock representations. However,
this does not preclude that both representations can be equivalent, as is
indeed the case for certain representations constructed by Kraus, Polley and
Reents \cite{KPR}. Therefore these so-called KPR representations are
interpreted as a background radiation field which is sufficiently chaotic in
the sense that the addition of bremsstrahlung only amounts to a slight
perturbation of them.

Motivated by this picture, D.~Buchholz has pointed out in \cite{Bu82} that in
front of such a background the space-like direction of the localisation cone
$\cC $ of $\psi_c$ should not be detectable any more. In mathematical terms,
the
representations $\pi_I$ and $\pi_I\circ\gamma_{\cC}$ (where $\pi_I$ corresponds
to
the KPR-like background) are expected to satisfy the so-called BF
(Buchholz-Fredenhagen) criterion, i.e.\ they should become unitarily
equivalent in restriction to the space-like complement of {\em any} space-like
cone $\cC_1$, in contrast to the situation prevailing in front of the vacuum.

Thus, trading the vacuum $\pi_0$ for a background $\pi_I$ is expected to lead
to an important improvement of the localisability properties of the charged
states and should therefore permit to carry through the BF variant
of the DHR analysis \cite{BF82,DoRo89} in this case too.

In this perspective, the aim of the present letter is to provide a consistency
check of these ideas for the case of (massless) theories with DHR-like
charges. We will show that in such theories, a DHR analysis can be based on a
wide class of background representations $\pi_I$ and that the main results of
this analysis are invariant with respect to the choice of that background.
More interestingly, also the sets of sectors satisfying Borchers'
criterion are invariant.

These results provide a justification for viewing such representations as a
background and corroborate the hope that, in theories with long-range
interactions, the superselection structure may indeed be described more easily
in relation to such a background rather than to the vacuum.

We end this introduction by explaining the main assumptions and some notation.
The starting point of the following analysis will be a quantum field
theory given in its vacuum representation. More specifically,
let  $(\cH_0,U_0,\Omega_0)$ be a vacuum Hilbert space, that is a
Hilbert space $\cH_0$ which carries a strongly continuous unitary
representation  $U_0$ of the translation group $\RR $ (of
$(1+s)$-dimensional Minkowski
space, also denoted by $\RR $) whose spectrum is contained in the
forward lightcone $\overline{\vlk} \subset \RR $ and such that the translation
invariant subspace of $\cH_0$ is spanned by a single unit vector $\Omega_0$.
Assume further that we are given a Haag-Kastler net
$$ \cO \longmapsto \cA(\cO) $$
of von Neumann algebras on $\cH_0$, i.e.\  an isotonous mapping from the
bounded regions $\cO\subset\RR $ to the von Neumann subalgebras of $\cB(\cH_0)$
which satisfies locality (i.e.\ $\cA(\cO_1) \subset \cA(\cO_2)'$ if
$\cO_1\subset{\cO_2}'$, in standard notation), translation covariance w.r.t.\
$\alpha_x \df {\rm Ad}U_0(x)$ (i.e.\ $\alpha_x(\cA(\cO))=\cA(\cO+x)$) and
weak additivity.
The quasilocal algebra  $\overline{\bigcup_{\cO}\cA(\cO)}^{\|\cdot\|}$
is also denoted by $\cA $ and is assumed to act irreducibly.
It will be of crucial importance later on that
one can conclude from these assumptions that the net $\cA $ has Borchers'
property B and that the C$^*$-algebra $\cA $ is simple,
cf.\ \cite{d'Ant}. (These conclusions do not, in fact, rely on the existence
and/or uniqueness of the vacuum vector $\Omega_0$.)

It is customary to introduce the following notations at this point:
the identical representation $\pi_0:\cA\longrightarrow\cB(\cH_0)$ is called the
vacuum
representation; it is the GNS representation of the vacuum state
$\omega_0=(\Omega_0,\cdot\: \Omega_0)$.
The spectral family associated with the translations $U_0$ is denoted by
$E_0$.

\section{Energy Components}
In this section, we will present certain folia of states, namely energy
components of positive energy representations. This notion has been introduced
by H.-J.\ Borchers and D.\ Buchholz  \cite{BuBo85} and discussed further by
R.\ Wanzenberg  \cite{Wa87}. We begin this section by recalling some
details which are relevant in the present context.

For the whole of this section, let $(\cH,\pi)$ be a positive energy
representation of $(\cA,\alpha)$. This means that there exists, on the Hilbert
space $\cH $, a strongly continuous representation $U_{\pi}$ of the
translations such that
$$ \pi\circ \alpha_x = {\rm Ad}U_{\pi}(x) \circ \pi \quad\mbox{and} \quad
   {\rm sp}U_{\pi}\subset \overline{\vlk}.    $$
It is then known \cite{Bo84} that the operators $U_{\pi}(x)$ can
be chosen such that $U_{\pi}(x)\in\pi(\cA)''$, and it is a remarkable fact
\cite{BuBo85} that there is a  unique choice for which,
on any subspace $\cH'\subset\cH $ which is invariant under $\pi(\cA)$,
the spectrum ${\rm sp}U_{\pi}|_{\cH'}$ has Lorentz-invariant lower
boundary. From now on, $U_{\pi}$ will always denote this canonical
implementation of the translations and $E_{\pi}$ the associated spectral
family.

The notion of energy contents of $\pi $-normal states can now be introduced.
For any compact set\footnote{In the sequel, $\Delta $
                              {\em always} stands for a compact subset
                              of $\overline{\vlk}$. Moreover, we will use the
                              notation $\Delta_q\df \overline{\vlk} \cap
                              (q-\overline{\vlk}) $ for the double cone in
                              momentum space with apices $0$ and $q\in\vlk$. }
$\Delta\subset\overline{\vlk}$, denote by
$$ \cS_{\pi}(\Delta) \df \Big\{ {\rm tr}\rho\pi(\cdot) \: \Big| \:
                                \rho\in\cI_1(\cH),\: \rho\geq 0,\:  {\rm
tr}\rho=1,\:
                                \rho E_{\pi}(\Delta)=\rho   \Big\}    $$
the set of all $\pi$-normal states which have energy-momentum in  $\Delta$.
Then $\cS_{\pi}=\overline{ \bigcup_\Delta \cS_{\pi}(\Delta)}^{\|\cdot\|} $ is
just the
folium of $\pi $. Another set $\tilde{\cS}_{\pi}$ of states, called the energy
component of $\pi $, is defined by
$$ \tilde{\cS}_{\pi} \df \overline{ \bigcup_{\Delta}
                           \tilde{\cS}_{\pi}(\Delta)}^{\|\cdot\|}     ,$$
where $\tilde{\cS}_{\pi}(\Delta)$ denotes the set of all locally $\pi $-normal
states in the weak closure of $\cS_{\pi}(\Delta)$. Of course,
$\cS_{\pi}\subset\tilde{\cS}_{\pi}$, and one can show that $\tilde{\cS}_{\pi}$
is a folium, too. Physically, it is interpreted as the set of states which
can be reached
from the folium of $\pi $ by operations requiring only a finite amount of
energy. Some important properties of the elements of  $\tilde{\cS}_{\pi}$ are
collected in the following Lemma, parts~2 and 3 of which are due to
R.\ Wanzenberg \cite{Wa87}.

\begin{lem}
Let $\omega\in\tilde{\cS}_{\pi}$ and denote by
$(\cH_{\omega},\pi_{\omega},\Omega_{\omega})$
its GNS triple.
\begin{enumerate}
\item Then $(\cH_{\omega},\pi_{\omega})$ is a locally $\pi $-normal positive
energy
      representation of $(\cA,\alpha)$.
\item Moreover, if $\omega\in\tilde{\cS}_{\pi}(\Delta_q)$ for some
      $q\in\overline{\vlk}$, then
      one has  $\Omega_{\omega}\in E_{\pi_{\omega}}(\Delta_q)\cH_{\omega}$.
\item In the situation of 2, one has

$\tilde{\cS}_{\pi_{\omega}}(\Delta)
\subset\tilde{\cS}_{\pi}(\Delta+\Delta_q-\Delta_q)$ for any $\Delta$.
\end{enumerate}
\end{lem}

Proof:
Using the positivity of the energy in the representation $\pi $,
part 1 follows from the fact that $\omega $ is locally $\pi $-normal
by arguments similar to those in \cite{BuDo84}.
(In \cite{BuDo84}, these arguments are
only applied to states $\omega\in\tilde{\cS}_{\pi}(\Delta_q)$,
but they carry over to norm limits of such states as well.)

For simplicity, part 2 will only be proved in the case that $\pi_{\omega}$ is
factorial. For the general case, see \cite{Wa87}. Let $\Delta_q$ be given. Then
the idea is to show that $\Omega_{\omega}$ does not have momentum outside
$\Delta_q$. To this end, fix some $p\in\overline{\vlk}\setminus\Delta_q$ and
choose
a neighbourhood $\cN_p\subset\overline{\vlk}\setminus\Delta_q$ of $p$ and an
open
set $\cN\subset {\rm sp}U_{\pi_{\omega}}$ such that
$$ (\Delta_q+\cN-\cN_p)\cap\overline{\vlk}=\emptyset.$$
(Such a choice is always possible because the lower boundary
of $ {\rm sp}U_{\pi_{\omega}}$ is
Lorentz invariant.) Now choose a test function $f$  satisfying
${\rm supp}\tilde{f}\subset\cN-\cN_p $ and take an arbitrary $A\in\cA $. Then,
$A(f)\df \int dx \,f(x)\alpha_x(A)$ is an element of $\cA $ and satisfies
$\pi(A(f)) E_{\pi}(\Delta_q)\cH_{\pi}=\{0\}$.
Since $\omega\in\tilde{\cS}_{\pi}(\Delta_q),$ this implies
$\omega(A(f)^*A(f))=0,$
hence $\pi_{\omega}(A(f))\Omega_{\omega}=0.$ This means that $\Omega_{\omega}$
is orthogonal to
$$ \cD \df {\rm span} \left\{ \pi_{\omega}(A(f))^*\Psi \Big|
\Psi\in\cH_{\omega},A\in\cA,
    {\rm supp}\tilde{f}\subset\cN-\cN_p \right\}.  $$
Since $\pi_{\omega}$ is factorial and $\cN-\cN_p$ is open, it follows by an
argument explained in \cite{BF82} (see, in particular, the proof of Prop.~2.2
therein)  that the closure of $\cD$ equals
$E_{\pi_{\omega}}({\rm sp}U_{\pi_{\omega}}+ \cN_p-\cN)\cH_{\omega}$. Thus,
$\Omega_{\omega}\in\cD^{\perp}$
yields
$$ \{ \Omega_{\omega} \}^{\perp} \supset \cD^{\perp\perp} = \overline{\cD} =
   E_{\pi_{\omega}}({\rm sp}U_{\pi_{\omega}}+ \cN_p-\cN )\cH_{\omega} \supset
   E_{\pi_{\omega}}(\cN_p)\cH_{\omega},    $$
where the last inclusion holds because ${\rm sp}U_{\pi_{\omega}}-\cN \ni 0 $.
{}From this, we get $E_{\pi_{\omega}}(\cN_p)\Omega_{\omega}=0$ or, as
$p\in\overline{\vlk}\setminus\Delta_q$ was arbitrary,
$\Omega_{\omega}\in E_{\pi_{\omega}}(\Delta_q)\cH_{\omega}.$

To prove part 3, let $\Psi\in E_{\pi_{\omega}}(\Delta)\cH_{\omega}$.
Part 2 and the cyclicity of $\Omega_{\omega}$ imply that
there exists in $\cA$  a sequence $(A_n)_{n\in\N}$  of
operators with energy-momentum support in $\Delta-\Delta_q$ such that
$\Psi=\lim_{n\to\infty} \pi_{\omega}(A_n)\Omega_{\omega}.$ From
$\omega\in\tilde{\cS}_{\pi}(\Delta_q)$
follows $\omega(A_n^*\cdot
A_n)\in\tilde{\cS}_{\pi}(\Delta_q+(\Delta-\Delta_q))=
\tilde{\cS}_{\pi}(\Delta+\Delta_q-\Delta_q)$, i.e.\
$$ \langle \Psi, \pi_{\omega}(\cdot) \Psi \rangle
   =\lim_{n\to\infty}\omega(A_n^*\cdot
A_n)\in\tilde{\cS}_{\pi}(\Delta+\Delta_q-\Delta_q).  $$
Thus  any vector state from $\cS_{\pi_{\omega}}(\Delta)$ lies in
$\tilde{\cS}_{\pi}(\Delta+\Delta_q-\Delta_q)$. One now gets the assertion  by
taking
convex combinations, norm limits and locally normal w$^*$-limits.
\Bix

The preceding lemma has shown that positivity of the energy is a property
which ``survives'' the process of going from the representation $\pi $ to
the GNS representation of a state in $\tilde{\cS}_{\pi}$. Other properties
survive as well, as for instance the compactness condition C$_{\sharp }$ of
Fredenhagen and Hertel \cite{FrHe,BuPo86} which can be formulated as follows:

{\bf Definition}: Condition C$_{\sharp }$ is said to be satisfied in the
(positive energy) representation $\pi $ if, for any $\Delta $ and any bounded
region $\cO\subset\RR $, the set
$$ \cS_{\pi}(\Delta)\mid_{\cA(\cO)} \equiv
   \left\{ \omega\mid_{\cA(\cO)} \Big|\, \omega\in\cS_{\pi}(\Delta) \right\}
$$
is contained in a  $\|\cdot\|$-compact subset of $\cA(\cO)^*$.

This condition controls the infrared properties of the model under
consideration, cf.\ \cite{BuPo86}. It has been established in the theory of a
massive free particle (in any space-time dimension) and in the theory of a
massless (scalar or vector) particle in at least $1+3$ space-time dimensions
\cite{BuJa, BuPo86} and is believed to hold in QED as well. In the present
context, it will play a technical role in the proof of Prop.~3.2 since it
allows (by part~1 of the following lemma) a simplification in the definition
of $\tilde{\cS}_{\pi}(\Delta)$ given above.

\begin{lem}
Let  C$_{\sharp }$ be satisfied in the representation $\pi $ and
let $\omega $ be a state in the weak closure of $\cS_{\pi}(\Delta)$. Then
\begin{enumerate}
\item $\omega$  is locally $\pi $-normal, i.e.\
$\omega\in\tilde{\cS}_{\pi}(\Delta)$;
\item C$_{\sharp }$ is satisfied in the GNS representation of $\omega $.

\end{enumerate}
\end{lem}

Proof:
Both parts follow from the fact  that, in restriction to
$\cA(\cO)$, any w$^*$-limit of $\cS_{\pi}(\Delta)$ is,
as a consequence of C$_{\sharp }$,
even a $\|\cdot\|$-limit of $\cS_{\pi}(\Delta)$. Part 1 now follows directly.
For part 2, we note that the above fact implies
$\tilde{\cS}_{\pi}(\Delta)\mid_{\cA(\cO)} \subset
\overline{\cS_{\pi}(\Delta)\mid_{\cA(\cO)}}^{\|\cdot\|} $, which, in view of
Lemma 2.1(3) yields the assertion.
\Bix

\section{Infravacuum Representations}
Intuitively, an infrared cloud (in QED) can be split, given any energy
threshold $\epsilon>0$, into a finite number $N_{\epsilon}$ of ``hard'' photons
and
infinitely many ``soft'' photons whose total energy is less than $\epsilon $.
As the creation of hard photons can be described by a quasilocal operation,
this picture motivates the following mathematical notion.

{\bf Definition}: An irreducible representation $(\cH_I,\pi_I)$ of $\cA $
is called an infravacuum representation if, for any $q\in\vlk $,
the set $\tilde{\cS}_{\pi_0}(\Delta_q)$ contains some $\pi_I$-normal  state.

With any arbitrarily small amount of energy, one can thus create from the
vacuum
representation some state in an infravacuum representation. In the terms of
\cite{Wa87}, this means that the transition energy between the sectors
$\pi_0$ and $\pi_I$ vanishes (cf.\ Prop.~3.2 below). As an example for
infravacuum
representations, we mention the  KPR representations,
i.e.\ a class of non-Fock representations
(of the free asymptotic electromagnetic field) devised by Kraus, Polley and
Reents \cite{KPR}, cf.\ also \cite{Hars}, so as to be stable (up to
unitary equivalence) under the bremsstrahlung produced in typical
collision processes of charged particles.
We note that with the above definition, the vacuum $\pi_0$  itself is an
infravacuum representation, which will be convenient in the sequel. However,
the notion of infravacuum representations is tailored to theories with
massless particles in the sense that in a purely massive theory, $\pi_0$
would be the only infravacuum representation.

The following lemma collects some basic properties of any infravacuum
representation.
\begin{lem} Let $\pi_I$ be an infravacuum representation. Then
\begin{enumerate}
  \item for any $q\in\vlk $,  $\tilde{\cS}_{\pi_0}(\Delta_q)$ contains a
        {\em pure} $\pi_I$-normal state;
  \item $\pi_I$ is a locally normal positive energy representation
        of $(\cA,\alpha)$;

  \item for any bounded $\cO $, the restriction of
        $\pi_I:\cA(\cO)\longrightarrow\pi_I(\cA(\cO))$
        to uniformly bounded subsets of $\cA(\cO)$ is a homoeomorphism
        with respect to the weak operator topologies.

\end{enumerate}
\end{lem}

Proof:
Choose, for some $q\in\vlk$, a state
$\omega\in\cS_{\pi_I}\cap\tilde{\cS}_{\pi_0}(\Delta_q)$ and let
$(\cH,\pi,\Omega)$
be its GNS triple. Since $\omega\in\cS_{\pi_I}$, $\pi$ is quasi-equivalent
to the irreducible representation $\pi_I$. This implies that $\pi(\cA)'$
is a type I factor  and thus contains some minimal projection $P$.
Obviously, $\pi|_{P\cH}\cong\pi_I$, which implies assertion 2 by
Lemma 2.1(1). Moreover, recalling $\Omega\in E_{\pi}(\Delta_q)\cH$, it also
implies that the pure
$\pi_I$-normal state  $\omega_P\df (P\Omega,\pi(\cdot)P\Omega)/\|P\Omega\|^2$
fulfils
$$ \omega_P\in \cS_{\pi_I}(\Delta_q)\subset \cS_{\pi}(\Delta_q)
   \subset \tilde{\cS}_{\pi}(\Delta_q)
   \subset\tilde{\cS}_{\pi_0}(2\Delta_q). $$
Since $q$ was arbitrary, this proves part 1.
Finally, part~3 follows by applying Corollary~7.1.16.\ of \cite{KaRi}
to the maps $\pi_I:\cA(\cO)\longrightarrow\pi_I(\cA(\cO))$ which are
isomorphisms
of von Neumann algebras since $\pi_I$ is faithful and locally normal.
\Bix

Let us remark that one can even establish local unitary equivalence between
$\pi_0$ and $\pi_I$  under the additional assumption that $\pi_I$ satisfies
weak additivity
\footnote{If additivity and not merely weak additivity is assumed for $\pi_0$,
additivity and hence weak additivity follow for $\pi_I$ by local normality.}.
The reasoning for this goes along the following line: According to
the arguments collected nicely in \cite{d'Ant}, there exist  Reeh-Schlieder
vectors in $\cH_0$ and $\cH_I$, i.e.\ the von Neumann algebras $\cA(\cO)$ and
$\pi_I(\cA(\cO))$ possess cyclic and separating vectors. From this, one
obtains  local unitary equivalence by applying Thm.~7.2.9.\ of \cite{KaRi}.
We recall, however, that an attempt to establish unitary implementability
of $\pi_I$ on an algebra pertaining to an unbounded region
(such as $\cO' $) would fail because $\pi_I$ need not be normal on such a
region. Indeed, the interesting case will be the  situation in which
$\pi_I$ does {\em not} fulfil the DHR criterion with respect to $\pi_0$.

To conclude this section about general properties of infravacuum
representations, we want to compare the nets $\tilde{\cS}_{\pi_0}$ and
$\tilde{\cS}_{\pi_I}$ of states.  In  the next proposition,
we will have to make the additional assumption that the vacuum state is unique
in the sense that, for some $q\in\vlk $, $\omega_0$ is the only vacuum state in
$\tilde{\cS}_{\pi_0}(\Delta_q)$.

\begin{prop}
Let $\pi_I$ be an infravacuum representation. Then we have, for any $\Delta $
and any $q\in\vlk $:
$$ \tilde{\cS}_{\pi_I}(\Delta) \subset
\tilde{\cS}_{\pi_0}(\Delta+\Delta_q-\Delta_q). $$
If, moreover, the defining representation $\pi_0$ satisfies   C$_{\sharp}$
and if the vacuum state is unique in the sense explained above,
then also the converse is true, namely
$$ \tilde{\cS}_{\pi_0}(\Delta) \subset
\tilde{\cS}_{\pi_I}(\Delta+\Delta_q-\Delta_q). $$
\end{prop}

Proof:
Using Lemma 3.1(1), the first statement follows immediately from
Lemma 2.1(3). To prove the second statement, we will first use the two
additional
assumptions to show $\omega_0\in \tilde{\cS}_{\pi_I}(\Delta_q)$ for any
$\Delta_q$. This is
done as follows.
Take some $\omega\in\cS_{\pi_I}\cap\tilde{\cS}_{\pi_0}(\Delta_q)$. It is easy
to see
that this implies $\omega\in\tilde{\cS}_{\pi_I}(\Delta_q)$. Now consider the
family
$(\omega_L)_{L>0}$ of states defined by
$\omega_L\df\frac{1}{|V_L|} \int_{V_L} d^{1+s}x\,\omega\circ\alpha_x$,
where $V_L\df [-L,L]^{(1+s)}\subset\RR $. Since
$\omega_L\in\tilde{\cS}_{\pi_I}(\Delta_q)$, any w$^*$-limit  $\tilde{\omega}$
of this
family is a  w$^*$-limit of $\tilde{\cS}_{\pi_I}(\Delta_q)$. Now $\pi_I$
satisfies
C$_{\sharp}$ by Lemma~2.2(2), so Lemma~2.2(1) can be applied to
$\tilde{\omega}$
and yields $\tilde{\omega}\in\tilde{\cS}_{\pi_I}(\Delta_q)$. In particular,
$\tilde{\omega}$ has positive energy. On the other hand,  $\tilde{\omega}$ is
translation invariant by construction, i.e.\ it is a vacuum state. Moreover,
by the first part of the present proposition,
$\tilde{\omega}\in\tilde{\cS}_{\pi_0}(2\Delta_q)$. Now if $q$ is sufficiently
small, the
uniqueness assumption yields $\tilde{\omega}=\omega_0$ whence
$\omega_0\in\tilde{\cS}_{\pi_I}(\Delta_q)$. Trivially, this conclusion remains
true
for any $q$. With this information, the second statement follows again from
Lemma~2.1(3).
\Bix

We remark, in addition, that the previous proposition does not, in general,
entail that one of the sets $\tilde{\cS}_{\pi_0}(\Delta)$ and
$\tilde{\cS}_{\pi_I}(\Delta)$ is contained in the other because the nets
$\Delta\longmapsto\tilde{\cS}_{\pi}(\Delta)$ ($\pi=\pi_0$ or $\pi_I$) need not
be regular from
the outside (which would mean $\tilde{\cS}_{\pi}(\Delta)= \,\,
\tilde{\!\!\tilde{\cS}}_{\pi}(\Delta)
\df\bigcap_{q}\tilde{\cS}_{\pi}(\Delta+\Delta_q)$),
but the outer regularized nets  $\,\,\tilde{\!\!\tilde{\cS}}_{\pi}$ coincide.
We note, as an aside, that the nets  $\cS_{\pi_0}$ and $\cS_{\pi_I}$, in
contrast, are regular from the outside, as follows from the continuity of the
spectral families $E_0$ and $E_{\pi_I}.$

\section{The DHR Criterion}
In the spirit of the theory of superselection sectors by Doplicher, Haag and
Roberts \cite{DHR}, it is  natural to consider, for any infravacuum
representation
$\pi_I$, the set of representations
$$ {\rm DHR}(\pi_I)\df \left\{ \pi\; \Big|\; \pi\mid_{\cA(\cO')} \cong
                       \pi_I\mid_{\cA(\cO')},\; \forall \cO\in\cK \right\}  $$
(where $\cK $ denotes the set of all double cones in $\RR $). The well-known
DHR analysis can be carried through under one additional assumption,
namely that of Haag duality. Thus, we will restrict our attention in the sequel
to those infravacuum representations which have this property, i.e.\
$$ \pi_I(\cA(\cO'))' = \pi_I(\cA(\cO)),\quad \forall \cO\in\cK.$$
We mention as an aside that Haag duality has been established for KPR
representations (cf.\ Sect.~3) if the net of observables fulfils an additional
condition corresponding to an algebraic version of Gauss' law \cite{Hars,
Drie}.

Instead of $ {\rm DHR}(\pi_I)$, one can now  study the set
$$ \Delta_{\pi_I,t} \df \bigcup_{\cO\in\cK}\Delta_{\pi_I,t}(\cO)  $$
of transportable localised endomorphisms of the C$^*$-algebra $\pi_I(\cA)$,
where $\Delta_{\pi_I,t}(\cO)$ denotes the subset of all such endomorphisms
which
act trivially on the subalgebra $\pi_I(\cA(\cO'))$. Transportability means
that, for any $x\in\RR $ the endomorphisms  $\rho_I $ and $\rho_{I,x}$ are
unitarily equivalent
in $\cH_{\pi_I}$, where $\rho_{I,x}$ is given by the action of the translations
$$\Delta_{\pi_I,t}(\cO) \longrightarrow \Delta_{\pi_I,t}(\cO+x): \rho_I
\longmapsto \rho_{I,x} \df
   {\rm Ad}U_{\pi_I}(x)\circ \rho_I\circ {\rm Ad}U_{\pi_I}(x)^*.       $$
We recall that $\Delta_{\pi_I,t}$ can be viewed as the set of objects of a
tensor
C$^*$-category\footnote{Viewing the index $I$ as a label which distinguishes
                    different infravacuum representations, we let for clarity
                    all objects of $\Delta_{\pi_I,t}$ carry an index $I$. (This
                    accuracy will not be necessary for the morphisms, however.)
                    In particular, $I$ may take on the value 0, then referring
                    to the vacuum.}
with subobjects and finite direct sums  \cite{DoRo}, also denoted
by $\Delta_{\pi_I,t}$, whose morphisms $T\in I_{\pi_I}(\sigma_I,\rho_I)$ from
$\sigma_I $ to $\rho_I $
are the intertwining operators $T\in\cB(\cH_{\pi_I})$ from $\sigma_I\circ\pi_I$
to
$\rho_I\circ\pi_I$. It is a crucial consequence of Haag duality that the
intertwiners are again local observables:
$$I_{\pi_I}(\sigma_I,\rho_I)\subset\pi_I(\cA(\cO))
    \quad\mbox{if} \quad\sigma_I,\rho_I\in\Delta_{\pi_I,t}(\cO) . $$

One  eventually has to pass from $ \Delta_{\pi_I,t}$ to the full subcategory
$\Delta_{\pi_I,f}$
of objects with finite statistical dimension, which is a tensor C$^*$-
category with conjugates. Its importance resides in the fact that it is
possible to recover from it (by a deep result of Doplicher and Roberts
\cite{DoRo89})
a compact gauge group and a field net with normal commutation
relations whose gauge invariant part contains all
finite-dimensional DHR sectors of $\pi_I(\cA)$.

It is important to show that the above-mentioned notions do not, in fact,
depend on the choice of the infravacuum representation $\pi_I$. Indeed, we
obtain the following slightly more general result:
\begin{prop}
Let $\pi_I$ be an irreducible, locally normal representation of $\cA $ and
assume that $\pi_0$ and $\pi_I$  satisfy Haag duality. Then, a bijective
functor
$$ F:  \Delta_{\pi_0,t}\longrightarrow  \Delta_{\pi_I,t}$$
is defined by the actions (on objects resp.\ on morphisms)
$$\begin{array}{ll}
  F(\rho_0)\:\df\, \pi_I\circ\pi_0^{-1}\circ\rho_0\circ\pi_0\circ\pi_I^{-1}, &
  \rho_0\in \Delta_{\pi_0,t} \\
  F(T)\df\, \pi_I\circ\pi_0^{-1}(T), & T\in I_{\pi_0}.
\end{array} $$
This functor restricts to an isomorphism
$F:\Delta_{\pi_0,f}\longrightarrow\Delta_{\pi_I,f}$ of
the corresponding tensor C$^*$-categories with conjugates.
\end{prop}

Proof:
As the algebra $\cA $ is simple, any representation is faithful, so
$\pi_I\circ\pi_0^{-1}: \pi_0(\cA)\longrightarrow\pi_I(\cA)$ is bijective.
Disregarding
for a moment the question of transportability, this means that
$F$, as defined above, is well-defined on the objects $\rho_0 $  and
remembering
$I_{\pi_0}\subset \pi_0(\cA)$, also on the morphisms. Obviously,
$F$ is bijective. One checks that $F(\rho_0)$ is localised in $\cO $ iff
$\rho_0 $ is,
which implies that $F(\rho_0)$ is transportable iff $\rho_0 $ is. Since the
statistical dimension of an object is a purely algebraic notion
\cite{DoRo95}, $F$ clearly restricts to
the subcategories of finite statistical dimension. As any bijective functor,
$F$ preserves all algebraic structures involved, such as subobjects,
direct sums, conjugates and the symmetries.
\Bix

Obviously, what lies at the heart of this proof is that $\pi_I\circ\pi_0^{-1}$
is a net isomorphism. The additional property of this isomorphism of being
bicontinuous (w.r.t.\ the weak operator topologies) in restriction to
norm-bounded subsets of local algebras (cf.\ Lemma~3.1.3) has not been
exploited yet but will play an  essential role in Section~6.

We should emphasise that the above Proposition and its proof
remain valid if  $\pi_0$  also is replaced by an  irreducible, locally normal
representation of $\cA $
which satisfies Haag duality. In particular, the spectral properties of this
representation do not enter into the argument.

As straightforward as the foregoing proposition may be, as important is its
interpretation: it shows that replacing the vacuum $\pi_0$ by an
infravacuum $\pi_I$ does not affect the superselection structure
of the theory. Not only the charges of the theory remain the same, but also
their fusion structure, their statistics, their localisation and their
transportability properties. In more pictorial terms, an infravacuum is suited
as well as the ``empty'' vacuum  as a background in front of which DHR theory
can take place.

\section{Cocycles and Their Spectral Properties}
We have seen in the previous section that the DHR superselection structure of
a theory does not depend on the infravacuum with respect to which it is
defined. In elementary particle physics, physically meaningful (DHR)
representations should, however, also fulfil what is called Borchers'
criterion, i.e.\ have positive energy. In the next two sections, we want to
convince ourselves that this additional property too is independent of the
infravacuum representation chosen.

It is well-known \cite{DHR} that Borchers' criterion
is fulfilled for all $\rho_0\in\Delta_{\pi_0,f}$ since the
subcategory of finite-dimensional objects is closed under conjugates. However,
it is important to note that the proof of this theorem,
in relying on the additivity
of the energy, makes crucial use of the fact that $\pi_0$ is a vacuum
representation and thus fails when $\pi_0$ is replaced by $\pi_I$.
Yet, any $\rho_I\in\Delta_{\pi_I,f}$
actually {\em does} fulfil Borchers' criterion,  as follows
from Prop.~6.3 below.

Leaving DHR theory for a moment, we first discuss a mathematical notion whose
relevance will soon become clear. Thus, let  $(\cH_I,U_I)$
be a pair  consisting of a Hilbert
space $\cH_I$ and a strongly continuous unitary  representation $U_I$ of $\RR $
whose spectral family will be denoted by $E_I$. Note that no condition is
imposed on its spectrum at this stage.

{\bf Definition}: A strongly continuous function
$\Gamma:\RR\longrightarrow\cU(\cH_I)$ is
called a cocycle over $(\cH_I,U_I)$ if it fulfils the cocycle equation
$$ \Gamma(x)\;\mbox{Ad}U_I(x)(\Gamma(y))=\Gamma(x+y) .$$
A cocycle is called $C$-spectral, where $C$ is some closed subset of $\RR $,
if it complies with the following condition: for any compact $\Delta\subset\RR
$,
one has
$$ \mbox{\rm sp}\big(\Gamma(\cdot)E_I(\Delta)\big) \subset C-\Delta.$$
Here, the spectrum $\mbox{\rm sp}A$ of a uniformly bounded, strongly continuous
operator-valued function $A:\RR\longrightarrow\cB(\cH_I)$ is, by definition,
the support
of its Fourier transform in the sense of operator-valued distributions. In
particular, one has $\int dx\,f(x)A(x)=0$ for any test function $f$ with
${\rm supp}\tilde{f}\cap{\rm sp}A=\emptyset $. Of course, if
$A=U:\RR\longrightarrow\cU(\cH_I)$ is a group homomorphism, ${\rm sp}A={\rm
sp}U$ coincides
with the common spectrum of the generators of the group.

The notion of spectral cocycles does not seem to have appeared in the
literature.
The idea of applying it to the above-mentioned problem and the main part of
Prop.~5.2 are due to D.~Buchholz.

The following basic properties of the spectrum will be important in the sequel:
\begin{lem}
Let $A,A_1$ and $A_2$ be uniformly bounded, strongly continuous
operator-valued function on $\RR$. Then, one has
\begin{enumerate}
\item ${\rm sp}A^*=-{\rm sp}A$, where $A^*(x)=(A(x))^*$;
\item ${\rm sp}(A_1+A_2)\subset{\rm sp}A_1\cup {\rm sp}A_2$;
\item ${\rm sp}(A_1 A_2)\subset{\rm sp}A_1+{\rm sp}A_2\;$ if $\;{\rm sp}A_1$ or
$\;{\rm sp}A_2$ is bounded.
\end{enumerate}
\end{lem}

Proof:
Parts 1 and 2  are elementary, whereas part 3 can be obtained,
by considering matrix elements, from the corresponding property of
scalar-valued
functions which in turn is the contents of \S\S1,2 and Th\'eor\`eme II
of \cite{Schwartz}.
\Bix

\begin{prop}
A bijective correspondence between strongly continuous unitary representations
$V:\RR\longrightarrow\cU(\cH_I)$ and cocycles $\Gamma $ is given by
$$\Gamma(x)=V(x)U_I(x)^*.$$
Moreover,  ${\rm sp}V\subset C$ iff $\:\Gamma $ is $C$-spectral.
\end{prop}

Proof:
The first assertion is obvious. As to the second one, let $V$ satisfy
${\rm sp}V\subset C$. Making use of the previous Lemma, one then has, for any
compact $\Delta\subset\RR$,
$$ {\rm sp}(\Gamma(\cdot)E_I(\Delta)) = {\rm
sp}(V(\cdot)\,U_I^*(\cdot)E_I(\Delta)) \subset
   {\rm sp}V-{\rm sp}(E_I(\Delta)U_I(\cdot))    \subset C-\Delta $$
which shows that $\Gamma $ is $C$-spectral. To prove the converse, let $\Gamma$
be
$C$-spectral. Let $\Delta\subset\RR $ be compact and choose a cover of $\Delta$
by a finite
 number of pairwise disjoint Borel sets $(\Delta_j)_{j=1,\dots,N}$. Then, again
by
 Lemma~5.1, one gets
\begin{eqnarray*}
 &\!\!\!{\rm sp}\Big(V(\cdot)E_I(\Delta)\Big)\!\!\!
 & \subset {\rm sp}\Big(V(\cdot)\sum_j E_I(\Delta_j)\Big)
   =  {\rm sp}\sum_j
\Big(\Gamma(\cdot)E_I(\Delta_j)\,E_I(\Delta_j)U_I(\cdot)\Big)   \\
 & &\subset\bigcup_j\Big({\rm sp}(\Gamma(\cdot)E_I(\Delta_j))+{\rm
sp}(E_I(\Delta_j)U_I(\cdot))
                \Big)\\
 & &  \subset \bigcup_j\Big(C-\Delta_j+\Delta_j\Big) \quad =\quad
C+\bigcup_j\Big(\Delta_j-\Delta_j\Big).
\end{eqnarray*}
Since the cover may be chosen such that the  maximal diameter of the sets
$\Delta_j$ is arbitrarily small, this implies  ${\rm sp}(V(\cdot)E_I(\Delta))
\subset C$
for any compact $\Delta\subset\RR $, which, in view of
${\rm s-}\lim_{\Delta\nearrow\RR}E_I(\Delta)={\bf 1}_{\cH_I}$,
finally yields ${\rm sp}V(\cdot)\subset C$.
\Bix

\section{Borchers' Criterion and Infravacua}
Let us now return to DHR theory. If $\pi_I$ is an infravacuum representation
of $\cA $, the role of what was denoted by $(\cH_I,U_I)$ in the previous
section is
of course taken over by $(\cH_{\pi_I},U_{\pi_I})$. For any object $\rho_I\in
\Delta_{\pi_I,t}$,
the following  set $Z^{\rho_I}$ of cocycles is of interest:

$$ Z^{\rho_I}\df\Big\{\Gamma^{\rho_I}\Big|\Gamma^{\rho_I}\;\mbox{is a cocycle
over}\;
  (\cH_{\pi_I},U_{\pi_I}) \;\mbox{and}\; \Gamma^{\rho_I}(x) \in
I_{\pi_I}(\rho_{I,x},\rho_I)
   \Big\} . $$
Calling an object $\rho_I $ covariant if there exists a strongly continuous
unitary
representation\footnote{Such a representation will be called an implementation
                     and denoted with the symbol $V_{\rho_I}$ in order to avoid
                     confusion with the unique canonical implementation
$U_{\pi}$
                     introduced in Sec.~2 for the positive energy
                     representations $\pi $ of $\cA $.}
$V_{\rho_I}:\RR\longrightarrow\cU(\cH_{\pi_I})$ which implements
the translations in  the representation $\rho_I\circ\pi_I$, we easily obtain
the following result by relating $V_{\rho_I}$ and $\Gamma^{\rho_I}$ as in
Prop.~5.2:
\begin{lem}
$Z^{\rho_I}$ is nonempty iff $\rho_I\in \Delta_{\pi_I,t}$ is covariant.
\end{lem}
In the rest of this section, we will study the behaviour of the cocycles
introduced above under the bifunctor $
F:\Delta_{\pi_0,t}\longrightarrow\Delta_{\pi_I,t}$ described in
Prop.~4.1. We now have cocycles $\Gamma^{\rho_0} \in Z^{\rho_0}$ and
$\Gamma^{\rho_I}\in Z^{\rho_I}$ which take on values in the groups of
local and unitary elements of $\pi_0(\cA)$ resp.\ $\pi_I(\cA)$. As a
consequence, not only they are in the domain of $F$ (which equals
$\pi_I\circ\pi_0^{-1}$ on intertwiners) resp.\ $F^{-1}$, but we even have
because of the continuity properties of $F$ established in Lemma~3.1(3)
(whose proof only relies on local normality --- cf.\ also the remark after
Proposition~4.1.):
\begin{lem}
Let $\rho_0\in \Delta_{\pi_0,t}$, $\rho_I\df F(\rho_0)$. Then  the functor $F$
maps $Z^{\rho_0}$ onto $Z^{\rho_I}$.
\end{lem}
In other words, $\rho_0 $ is covariant iff $\rho_I$
is\footnote{We will stick to the notation $\rho_I\df F(\rho_0)$ in the
sequel.}. Moreover,
an elementary calculation shows that, in this case,
implementations $V_{\rho_0}$ resp.\  $V_{\rho_I}$ of the translations
for $\rho_0\circ\pi_0$ resp.\  $\rho_I\circ\pi_I$ can be obtained from each
other by the
formula
\begin{equation} \label{F1}
 V_{\rho_I}(x) =
 \pi_I\circ\pi_0^{-1} \Big( V_{\rho_0}(x)\,U_{\pi_0}(x)^* \Big) \,U_{\pi_I}(x).
\end{equation}

We will now investigate in what extent the functor $F$ also respects spectral
properties. The inclusions obtained in Prop.~3.2 are taken as a starting point,
but we will comment on more general situations at the end of this section.
\begin{prop}
Let $\pi_I$ be an irreducible, locally normal positive energy representation
of $\cA $ and assume that $\pi_0$ and $\pi_I$  satisfy Haag duality. Denote by
$$ F:  \Delta_{\pi_0,t}\longrightarrow  \Delta_{\pi_I,t}$$ the bifunctor of
Prop.~4.1. If both inclusions
of Prop.~3.2 are valid, then one obtains for any covariant
object $\rho_0\in\Delta_{\pi_0,t}$:

\begin{enumerate}
  \item $F$ maps the $C$-spectral cocycles in $Z^{\rho_0}$ onto the
        $C$-spectral ones in $Z^{\rho_I}$.
  \item If $V_{\rho_0}$ and $V_{\rho_I}$ are related by (\ref{F1})
        then ${\rm sp}V_{\rho_0} ={\rm sp}V_{\rho_I}$.
  \item $\rho_0 $ has positive energy iff $\rho_I$ has.
  \item If $\rho_0 $ and $\rho_I$ have positive energy and are finite direct
sums
        of irreducibles, then the respective canonical implementations
        $U_{\rho_0\circ\pi_0}$ and $U_{\rho_I\circ\pi_I}$ of the translations
        are related by (\ref{F1}); in particular, their spectra coincide:
        $${\rm sp}U_{\rho_0\circ\pi_0}={\rm sp}U_{\rho_I\circ\pi_I}.$$
\end{enumerate}
\end{prop}

Proof:
1. Let $\Gamma^{\rho_0}\in Z^{\rho_0}$ be $C$-spectral and let
$\Gamma^{\rho_I}(x)\df F(\Gamma^{\rho_0}(x))$. It will be shown in the next
lemma
that this entails, for any compact set $\Delta$,
$$ \mbox{\rm sp}\Big(\Gamma^{\rho_I}(\cdot)E_I(\Delta)\Big) \subset
\bigcap_{\Delta_0} \big\{
    C-\Delta_0 \mid \tilde{\cS}_{\pi_I}(\Delta) \subset
\tilde{\cS}_{\pi_0}(\Delta_0) \big\}.$$
As a consequence of the first inclusion of Prop.~3.2, $\Delta_0$ can be chosen
to be any neighbourhood of $\Delta $. This implies that the right hand side
reduces to $C-\Delta $, i.e.\ $\Gamma^{\rho_I}$ is $C$-spectral. The converse
is
proved along the same lines, now using the second inclusion of Prop.~3.2.

2. Due to  (\ref{F1}), part~1 applies to the
cocycles $\Gamma^{\rho_0}(x)\df V_{\rho_0}(x)U_0(x)^*$
and $\Gamma^{\rho_I}(x)\df V_{\rho_I}(x)U_I(x)^*$.
Using Prop.~5.2, this yields for any closed $C\subset\RR:$
$${\rm sp}V_{\rho_0}\subset C\quad\mbox{iff}\quad {\rm sp}V_{\rho_I} \subset
C,$$
which immediately gives the assertion. From this, part~3 follows directly.

As to part~4, we first consider the special case when $\rho_0 $
(and therefore also $\rho_I$) is irreducible. Then all implementations of the
translations coincide up to a  one-dimensional one, i.e.\ a phase factor.
Together with  part~2, this yields  the following equivalences:
$$V_{\rho_0}=U_{\rho_0\circ\pi_0}
  \Leftrightarrow {\rm sp}V_{\rho_0}\; \mbox{has L\,i\,l\,b}
  \Leftrightarrow {\rm sp}V_{\rho_I}\; \mbox{has L\,i\,l\,b}
  \Leftrightarrow V_{\rho_I}=U_{\rho_I\circ\pi_I}              $$
(where ``L\,i\,l\,b'' means ``Lorentz-invariant lower boundary''). Applying
part~2 again gives all of the assertion in this case. In the general case of
reducible $\rho_0 $, we prove
$V_{\rho_0}=U_{\rho_0\circ\pi_0} \Leftrightarrow
V_{\rho_I}=U_{\rho_I\circ\pi_I}$
as follows: if $W_j\in I_{\pi_0}(\rho_{0,j},\rho_0)$
denote intertwiners which perform the decomposition of $\rho_0 $ into
irreducibles $\rho_{0,j}$, then $F(W_j)$ decompose $\rho_I$ into irreducibles
$\rho_{I,j}=F(\rho_{0,j})$. The implementations $V_{\rho_0}$ and $V_{\rho_I}$
decompose into subrepresentations $W_j^*V_{\rho_0}(x)W_j$ and
$F(W_j^*)V_{\rho_I}(x)F(W_j)$ on the corresponding subspaces and these
subrepresentations can be seen to be related by (\ref{F1}) iff $V_{\rho_0}$
and $V_{\rho_I}$ are. Since the latter are, by definition, canonical
iff all their subrepresentations are, the assertion follows from the special
case discussed first.
\Bix

\begin{lem}
Let $\Gamma^{\rho_0}$ be $C$-spectral. Then we have for any compact $\Delta$:
$$ \mbox{\rm sp}\Big(\Gamma^{\rho_I}(\cdot)E_I(\Delta)\Big) \subset
\bigcap_{\Delta_0} \big\{
  C-\Delta_0 \mid \tilde{\cS}_{\pi_I}(\Delta) \subset
\tilde{\cS}_{\pi_0}(\Delta_0) \big\}.$$
\end{lem}

Proof:
Let $\Delta_0$ be such that
$\tilde{\cS}_{\pi_I}(\Delta) \subset \tilde{\cS}_{\pi_0}(\Delta_0)$ and let $f$
be a
test function satisfying ${\rm supp}\tilde{f}\cap(C-\Delta_0)=\emptyset $.
Setting
$ \Gamma^{\rho_0}(f) \df \int dx\,f(x)\Gamma^{\rho_0}(x)$,
we obtain $\Gamma^{\rho_0}(f) E_0(\Delta_0)=\int
dx\,f(x)\Gamma^{\rho_0}(x)E_0(\Delta_0)=0$ since
$\Gamma^{\rho_0}$ is $C$-spectral. This means
$\omega(\pi_0^{-1}(\Gamma^{\rho_0}(f)^*\Gamma^{\rho_0}(f)))=0$ for any vector
state
$\omega\in\cS_{\pi_0}(\Delta_0)$, hence (by weak continuity) for any
$\omega\in\tilde{\cS}_{\pi_0}(\Delta_0)$ and in particular for any
$\omega\in\tilde{\cS}_{\pi_I}(\Delta)$. We thus have
$\pi_I\circ\pi_0^{-1}(\Gamma^{\rho_0}(f))E_I(\Delta)=0$ and hence by local
normality $\int
dx\,f(x)(\pi_I\circ\pi_0^{-1}(\Gamma^{\rho_0}(x))E_I(\Delta)=0$.
(For this argument, one has to realize that $f$ can be approximated
in $L^1(\RR)$ by test functions with compact support.) As this holds
for any $f$ with ${\rm supp}\tilde{f}\cap(C-\Delta_0)=\emptyset $,
it follows that
${\rm sp}(\pi_I\circ\pi_0^{-1}(\Gamma^{\rho_0}(\cdot))E_I(\Delta))\subset
C-\Delta_0$. This
yields the assertion.
\Bix

The inclusions of Prop.~3.2 played a crucial role in the previous proposition.
However, partial results remain valid if less information about the relation
between $\tilde{\cS}_{\pi_I}$ and $\tilde{\cS}_{\pi_0}$ is available. For
instance, let us merely assume (instead of these inclusions) that
there exists a compact neighbourhood $\cN $
of $0$ such that (cf.\ Lemma~2.1(3))
$$\tilde{\cS}_{\pi_I}(\Delta)\subset\tilde{\cS}_{\pi_0}(\Delta+\cN)\quad
  \mbox{for all compact sets $\Delta$.}  $$
In this case, we obtain with the same arguments as above that the cocycle
$\Gamma^{\rho_I} \df F(\Gamma^{\rho_0}(\cdot))$  is $(C-\cN)$-spectral if
$\Gamma^{\rho_0} $
is $C$-spectral. In terms of  $V_{\rho_0}$  and $V_{\rho_I}$,
this means
$${\rm sp}V_{\rho_I} \subset{\rm sp}V_{\rho_0}-\cN.$$
In general, all detailed information on the shape of
${\rm sp}V_{\rho_I}$ (such as the size of possible mass gaps) is lost,
but what can easily be seen is that $\rho_I$ has positive energy if $\rho_0 $
has.
The canonical implementations of the translations need, in
general, not be related by (\ref{F1}) any more. A similar reasoning applies
if the roles of  $\pi_0$ and $\pi_I$ are exchanged.

\section{Conclusion and Outlook}
We have seen that, as far as DHR theory is concerned, the role of the vacuum
representation $\pi_0$ can be taken over by any locally normal representation
$\pi_I$ satisfying Haag duality. Moreover, we have given sufficient
conditions on $\pi_I$ which ensure that the class of representations
fulfilling Borchers' criterion is independent of $\pi_I$. These
representations may be interpreted as background fields whose interaction
with the charged particles of the model is weak.

Natural candidates for states describing such a background were elements of
the energy component of the vacuum. The latter notion, introduced in
\cite{BuBo85}, deserves some more interest on its own, and one might ask  in
the spirit of Section~3 under which circumstances other properties such as,
e.g.,  Haag duality carry over from $\pi_0$ to $\pi_I$.

It seems thus that for most purposes, the characteristic feature of a vacuum
representation, namely the existence of a translation invariant vector in
the Hilbert space of that representation, is not an essential property.
However, we recall that this property plays a crucial role for proving that
the energy-momentum spectra are additive under the fusion of covariant
sectors:
$$\mbox{\rm sp}U_{\rho_1\circ\rho_2\circ\pi_0} \supset \mbox{\rm
sp}U_{\rho_1\circ\pi_0}
    + \mbox{\rm sp}U_{\rho_2\circ\pi_0}   $$
Trivially, using the results of Prop.~6.3, this yields
$$\mbox{\rm sp}U_{F(\rho_1\circ\rho_2)\circ\pi_I} \supset
  \mbox{\rm sp}U_{F(\rho_1)\circ\pi_I}    + \mbox{\rm
sp}U_{F(\rho_2)\circ\pi_I},   $$
but it would of course be interesting to know under which assumptions the
latter inclusion can be derived without relying on the vacuum sector.

Finally, returning to Buchholz' proposal of reconciling QED with
superselection theory, several interesting questions arise. First, it has to
be clarified which parts of the present work (in particular of Sections~4
and 6) can be generalised from DHR-like localised charges to charges
localised in space-like cones. Second, it will be important to check in
specific models whether there exist backgrounds which permit better
localisation properties of charges of electric type than the vacuum does.
Candidates for such models might be the one proposed by Herdegen
\cite{Herdegen} or a simple model recently introduced by Buchholz et al.\
in \cite{BDMRS}.

\vspace*{1em}
{\bf Acknowledgements:}
I am deeply indebted to Prof.\ D.\ Buchholz for
numerous helpful discussions and constant interest in this
work. Financial support from the Deutsche Forschungsgemeinschaft
(``Graduiertenkolleg Theoretische Elementarteilchenphysik'') is also
gratefully acknowledged.

\end{document}